\documentclass[final,3p,times]{elsarticle}
\usepackage{amssymb,bbold}
\usepackage{amsmath}
\usepackage[utf8x]{inputenc}
\usepackage{natbib}
\usepackage{hyperref}

\hypersetup{
  colorlinks=true,linkcolor=blue,citecolor=blue,
  filecolor=blue,urlcolor=blue,breaklinks=true
}
\biboptions{sort&compress}
\journal{Annals of Physics}
\RequirePackage{color}

\begin{document}

\begin{frontmatter}

\title{Quantum motion of a point particle in the presence of the Aharonov--Bohm potential in curved space}

\author[ufma]{Edilberto O. Silva}
\ead{edilbertoo@gmail.com}

\author[unb]{S\'{e}rgio C. Ulhoa}
\ead{sc.ulhoa@gmail.com}

\author[ucl,uepg]{Fabiano M. Andrade}
\ead{f.andrade@ucl.ac.uk,fmandrade@uepg.br}

\author[ufcg,ufla]{Cleverson Filgueiras}
\ead{cleversonfilgueiras@yahoo.com.br}

\author[unb,unbg]{R. G. G. Amorin}
\ead{ronniamorim@gmail.com}

\address[ufma]{
  Departamento de F\'{i}sica,
  Universidade Federal do Maranh\~{a}o,
  Campus Universit\'{a}rio do Bacanga,
  65085-580 S\~{a}o Lu\'{i}s, Maranh\~{a}o, Brazil
}
\address[unb]{
  Instituto de F\'{i}sica,
  Universidade de Bras\'{i}lia,
  70910-900, Bras\'{i}lia, Distrito Federal, Brazil
}
\address[ucl]{
 Department of Computer Science,
 University College London,
 WC1E 6BT London, United Kingdom
}
\address[uepg]{
  Departamento de Matem\'{a}tica e Estat\'{i}stica,
  Universidade Estadual de Ponta Grossa,
  84030-900 Ponta Grossa, Paran\'{a}, Brazil
}
\address[ufcg]{
  Departamento de F\'{i}sica,
  Universidade Federal de Campina Grande,
  Caixa Postal 10071,
  58109-970, Campina Grande, Para\'{i}ba, Brazil
}
\address[ufla]{Departamento de Física (DFI),
Universidade Federal de Lavras (UFLA), Caixa Postal 3037,
37200-000, Lavras, Minas Gerais, Brazil}

\address[unbg]{
  Faculdade Gama,
  Universidade de Bras\'{i}lia,
  Setor Leste (Gama),
  72444-240, Bras\'{i}lia, Distrito Federal, Brazil
}

\begin{abstract}
The nonrelativistic quantum dynamics of a spinless charged particle in the presence of the Aharonov--Bohm potential in curved space is considered.
We chose the surface as being a cone defined by a line element in polar coordinates.
The geometry of this line element establishes that the motion of the particle can occur on the surface of a cone or an anti--cone.
As a consequence of the nontrivial topology of the cone and also because of two--dimensional confinement, the geometric potential should be taken into account.
At first, we establish the
conditions for the particle describing a circular path in such a context.
Because of the presence of the geometric potential, which contains a
singular term, we use the self--adjoint extension method in order to describe
the dynamics in all space including the singularity.
Expressions are
obtained for the bound state energies and wave functions.
\end{abstract}

\begin{keyword}
  Self--adjoint extension \sep
  Aharonov-Bohm problem \sep
  Geometric potential \sep
  Bound state
\end{keyword}

\end{frontmatter}

\section{Introduction}

\label{sec:introduction}

Over the years, condensed matter systems have been a proper environment to
test theoretical physics. Advanced ideas such as the concept of torsion in
geometry have been used in this context. For instance Cosserat continuum
behaves as if it was a Riemann--Cartan manifold endowed with a torsion
tensor \cite{Hehl:2007bn}. Also it is well known that a particle moving in a
solid with topological defects is analogous to a 3D gravity model with
torsion \cite{AoP.1992.216.1,arXiv:cond-mat/0407469}. In addition, fluid
systems have been used to verify Hawking radiation which otherwise would be
extremely difficult to study \cite{arXiv:1401.6612}.

Due to the great development of new materials such as graphene \cite
{Science.2004.306.666,NatureMat.2007.6.183,RMP.2009.81.109}, the use of
refined concepts like those in geometry became mandatory to properly describe
them. It is interesting to note that these theoretical ideas migrated to a
central role in the prediction of the behavior of such systems. This feature
is similar to what happened in General Relativity in which the dynamics of a
particle is defined by the space-time curvature \cite{carroll2013spacetime}.
Let us consider an electron moving on a 2D curved graphene surface, such as a
carbon nanotube, how does it affects its dynamics? In such a context it is
not possible to construct a Classical Lagrangian that reveals the proper
dynamics. It is necessary to take into account the curvature and the metric
tensor of the surface to which the electron is bounded.

To address the problem of a particle confined to a surface $S$ which is
embedded on a 3D space, at least two different approaches were developed.
The first one is based on a purely 3D geometry \cite{RevModPhys.29.377},
whilst the second one, proposed by da Costa \cite{PRA.1981.23.1982}, is
constructed as a limit of a 3D space to the curved surface $S$. This latter
process is equivalent to embedding the surface into an ordinary 3D Euclidean
space. As a consequence, the wave function splits into two parts, one of
them working as if there was an effective potential constructed in terms of
the mean and Gaussian curvatures. Thus, through this procedure, the particle
is subjected to the so called geometric potential \cite{PRA.1981.23.1982}.
Recently, the da Costa proposition appeared in various physical contexts as,
for instance, in the derivation of the Pauli equation for a charged spin
particle confined to move on a spatially curved surface in the presence of
an electromagnetic field \cite{PRA.2014.90.042117}, in the study of
curvature effects in thin magnetic shells \cite{PRL.2014.112.257203}, in
effects of non--zero curvature in a waveguide to investigate the appearance
of an attractive quantum potential which crucially affects the dynamics in
matter--wave circuits \cite{SR.2014.4.5274}, in the quantum mechanics of a
single particle constrained to move along an arbitrary smooth reference
curve by a confinement that is allowed to vary along the waveguide \cite
{PRA.2014.89.033630}, to derive the exact Hamiltonians for Rashba and cubic
Dresselhaus spin--orbit couplings on a curved surface with an arbitrary
shape \cite{PRB.2013.87.174413}, in the study of high--order--harmonic
generation in dimensionally reduced systems \cite{PRA.2013.88.033837}, to
explore the effects arising due to the coupling of the center of mass and
relative motion of two charged particles confined on an inhomogeneous helix
with a locally modified radius \cite{PRE.2013.88.043202}, to study the
dynamics of shape--preserving accelerating electromagnetic wave packets in
curved space \cite{PRX.2014.4.011038}, in the derivation of the Schr\"{o}
dinger equation for a spinless charged particle constrained to move on a
curved surface in the presence of an electric and magnetic field \cite
{PRL.2008.100.230403}, etc.

In this article, we intend to use the same set of electric and magnetic
fields of Ref. \cite{PRL.2008.100.230403} for a spinless charged particle,
but now confined to a conical surface and in a presence of an Aharonov-Bohm
potential \cite{PR.1959.115.485}. We analyze the case of a charged particle
describing a circular path and also the general dynamics in all space,
including the $r=0$ region. We use the self--adjoint extension method to
determine the most relevant physical quantities, such as energy spectrum and
wave functions by applying boundary conditions allowed by the system.

The paper is organized as follows. In section \ref{sec:EM}, we recall the
main ideas concerning the dynamics of a spinless charged particle on a
curved surface under the influence of an electromagnetic field. We start
with the Schr\"{o}dinger Hamiltonian and couple it to the curvature and the
electromagnetic potential. In section \ref{sec:CP}, we establish the
conditions for the particle to describe a circular path. We determine the
expressions for the energy eigenvalues, wave functions and discuss the role
played by the curvature on them. In section \ref{sec:SA}, we briefly discuss
some concepts of the self--adjoint extension method. We analyze the
particle's dynamics when it lies either on a cone or on an anti--cone. After
applying the boundary conditions allowed by the system, we obtain
expressions for the bound state energies and wave functions in both cases.
Finally, in section \ref{sec:conclusions}, we present our concluding
remarks. In this work we use units such as $\hbar=c=1$.

\section{Equations of motion}

\label{sec:EM}

In this section, we introduce the equations of motion. We consider the
refined fundamental framework for the thin--layer quantization scheme
discussed in \cite{arXiv:1502.02110v1}, where a sound performing sequence in
the thin--layer quantization process is addressed. The case we are dealing
with will coincide with that which comes from the semiclassical method
applied by da Costa to investigate the effective quantum dynamics for a
constrained particle \cite{PRA.1981.23.1982,PRL.2008.100.230403}. Thus, we
start with the Schr\"{o}dinger equation
\begin{equation}
H\psi =i\frac{\partial}{\partial t}\psi ,
\end{equation}
where the Hamiltonian is given by
\begin{equation}
H=\frac{\hat{p}_{\mu}\hat{p}^{\mu}}{2M}+V(q^{\mu}),
\end{equation}
where $\hat{p}^{\mu}=-i\nabla^{\mu}$ and the index $\mu $ runs from 1 to 3.
\ The coordinate $q_{3}$ is the one transverse to a thin interface. In the
limit $q_{3}\rightarrow 0$, the metric tensor confined on a surface $S$ is
given by \cite{arXiv:1502.02110v1}\
\begin{equation}
\tilde{g}_{\mu \nu}=\left(
\begin{array}{ccc}
g_{11} & g_{12} & 0 \\
g_{21} & g_{22} & 0 \\
0 & 0 & 1
\end{array}
\right),  \label{2Dmetric}
\end{equation}
and the first part of the above Hamiltonian reads as $\hat{p}_{\mu}\hat{p}
^{\mu}= \tilde{g}_{\mu \nu}\hat{p}^{\mu}\hat{p}^{\nu}$. Such a behavior
determines an immersion of a 2D geometry into a 3D Euclidean space. Thus, we
have a 2D effective metric $g_{ij}$. In Ref. \cite{PRA.1981.23.1982}, the
metric in Eq. (\ref{2Dmetric}) suggests a separable wave function in the
form $\psi (q^{1},q^{2},q^{3})=\psi_{s}(q^{1},q^{2})\psi_{N}(q^{3})$. This
splits the movement into two, one constrained on a surface and another one
which takes place on a normal direction of such a surface. In the normal
direction, the dynamics is governed by the usual Hamiltonian $
H_{N}=\partial_{3}\partial^{3}/2M+V_{\lambda}(q^{3})$, where the confining
potential $V_{\lambda}(q^{3})$ is assumed to localize the particle on the
surface $S$. However, a fundamental framework for the thin-layer
quantization scheme is not explicitly defined in it. Here we follow the
explicitly refined fundamental framework for the thin-layer quantization
scheme presented in \cite{arXiv:1502.02110v1}, where the limit $
q_{3}\rightarrow 0$ must be performed after calculating all curvilinear
coordinate derivatives. We also consider the minimal coupling with the
electromagnetic field by means of the prescription
\begin{equation}
\hat{p}^{\mu}\rightarrow \hat{p}^{\mu}-QA^{\mu},
\end{equation}
where $Q$ is the charge of the particle and $A^{\mu}$ is the potential
vector component. Therefore, the Schr\"{o}dinger equation that describes the
dynamics of a spinless charged particle bounded to a thin interface under
the effect of electric and magnetic fields is given by
\begin{align}
i\frac{\partial}{\partial t}\psi = {} &\frac{1}{2M}\left[ -\frac{1}{\sqrt{g}}
\partial_{i}\left(\sqrt{g}g^{ij}\partial_{j}\right) +\frac{iQ}{\sqrt{g}}
\partial_{i}\left( \sqrt{g}g^{ij}A_{j}\right) +2iQg^{ij}A_{i}\partial
_{j}+Q^{2}g^{ij}A_{i}A_{j}+V_{s}(\mathbf{r})+QV(\mathbf{r})\right] \psi
\notag \\
& - \frac{1}{2M}\left[ iQ\left( \partial_{3}A^{3}\right) +2iQA^{3}\partial
_{3}-Q^{2}\left( A^{3}\right)^{2}\right] \psi +\left[ -\frac{\partial
_{3}\partial^{3}}{2M}+V_{\lambda}(q^{3})\right] \psi,  \label{eqnew}
\end{align}
where $\mathbf{r}=\mathbf{r}(q_{1},q_{2})$, $V_{s}(\mathbf{r})$ is a
potential due to the geometry of the surface and $V(\mathbf{r})$ is the
electric potential on the surface. The Coulomb gauge sets $\mathbf{\nabla}
\cdot \mathbf{A}=0$ and in the picture described here it gives
\begin{equation}
-\frac{1}{\sqrt{g}}\partial_{i}\left( \sqrt{g}g^{ij}A_{j}\right)
+\partial_{3}A^{3}+2\mathcal{H}A^{3}=0,
\end{equation}
where $\mathcal{H}$ is the mean curvature. Notice that the Lorentz gauge and
the effective Schr\"{o}dinger equation cannot be decoupled from the mean
curvature of the surface simultaneously, except when $A^{3}=0$. By considering $
\psi(q^{1},q^{2},q^{3})=\psi_{s}(q^{1},q^{2})\psi_{N}(q^{3})$, we can write
\begin{equation}
i\frac{\partial}{\partial t}\psi_{N}= \left[ -\frac{\partial_{3}\partial^{3}
}{2M}+V_{\lambda}(q^{3}) \right] \psi_{N},  \label{q3}
\end{equation}
and
\begin{equation}
i\frac{\partial}{\partial t}\psi_{s}=\frac{1}{2M}\left[ -\frac{1}{\sqrt{g }}
\partial_{i}\left( \sqrt{g}g^{ij}\partial_{j}\right) +\frac{iQ}{\sqrt{g}}
\partial_{i}\left( \sqrt{g}g^{ij}A_{j}\right) +2iQg^{ij}A_{i}\partial
_{j}+Q^{2}g^{ij}A_{i}A_{j}+V_{s}(\mathbf{r})+QV(\mathbf{r})\right] \psi_{s},
\label{hspinless}
\end{equation}
which are the decoupled equations found by da Costa in a semi--classical
approach. These equations do not encompass the influence of the interface
thickness $d$ \cite{arXiv:1502.02110v1}. This case leads to energy shifts
that we ignore in this work.

Here, it is interesting to notice how the break in the isotropy of the space
arises. As a matter of fact, the velocity operators became noncommutative
usually due to the presence of the magnetic field. Since $M\hat{v}^{i}=\hat{p
}^{i}-QA^{i}$, we write the commutator
\begin{align}
\lbrack \hat{v}^{i},\hat{v}^{j}] = {} &\frac{iQ}{M^{2}}\left(
\nabla^{i}A^{j}-\nabla^{j}A^{i}\right),  \notag \\
= {} & \frac{iQ}{M^{2}}\left( g^{il}\partial_{l}A^{j}
+g^{il}\Gamma_{ml}^{j}A^{m}-g^{jl}\partial_{l}A^{i}
-g^{jl}\Gamma_{ml}^{i}A^{m}\right) ,  \notag \\
= {} & \frac{iQ}{M^{2}} \left(
F^{ij}+g^{il}\Gamma_{ml}^{j}A^{m}-g^{jl}\Gamma_{ml}^{i}A^{m} \right).
\end{align}
Therefore, we see that $\hat{v}^{1}$ does not commutes with $\hat{v}^{2}$,
not just because the existence of the magnetic field but also by the
geometry of the surface to which the movement is bounded.

\section{Motion in a Aharonov--Bohm potential}

Now, let us apply Eq. (\ref{hspinless}) to the Aharonov--Bohm problem. At
this point, we can make a connection with the description of continuous
distribution of dislocations and disclinations in the framework of
Riemann--Cartan geometry of Ref. \cite{AoP.1992.216.1}. If the particle is
now bounded to a surface with a disclination located in the $r=0$\ region,
the corresponding metric tensor, in cylindrical coordinates, is defined by
the line element (see Ref. \cite{JMP.2012.53.122106} for more details.)
\begin{equation}
ds^{2}=dr^{2}+\alpha^{2}r^{2}d\theta^{2},  \label{metric}
\end{equation}
with $0\leq r<\infty $, $0\leq \theta \leq 2\pi $, which describes a conical
surface. For $0<\alpha <1$ (a deficit angle), the metric (\ref{metric})
describes an actual cone, whilst for $\alpha=1$ the cone turns into a plane,
and for $\alpha >1$ (an excess angle), the resulting surface is an
anti--cone. It is worthwhile to observe that the line element in (\ref
{metric}) can be compared with the metric of the spacetime produced by a
thin, infinite, straight cosmic string (for the special case of $dt=dz=0$)
\cite{Book.2000.Vilenkin}
\begin{equation}
ds^{2}=-c^{2}dt^{2}+dr^{2}+\tilde{\alpha}^{2}r^{2}d\theta^{2}+dz^{2},
\label{m2}
\end{equation}
where the parameter $\tilde{\alpha}$ is given in terms of the linear mass
density of the cosmic string $\mu$ by $\tilde{\alpha}=1-4\mu/c^{2}$, which
is smaller than the unity. Moreover, the metric in (\ref{m2}) is locally
flat, but globally it is not \cite{Book.2000.Vilenkin,PRD.1981.23.852} and
through the metric (\ref{m2}), among others effects \cite
{PRD.1987.35.3779,JHEP.2004.2004.16,PRD.2012.85.045016,
EPJC.2014.74.3187,PRD.2012.85.041701}, we can study the gravitational analog
of the Aharonov--Bohm effect \cite
{NC.1967.52.129,JPA.1981.14.2353,PRD.1987.35.2031}.

It is known that the curvature tensor of the metric (\ref{metric}), when
considered as a distribution, is of the form \cite{SPD.1977.22.312}
\begin{equation}
R_{12}^{12}=R_{1}^{1}=R_{2}^{2}=2\pi \left( \frac{1-\alpha }{\alpha }\right)
\delta _{2}(r),  \label{tensorR}
\end{equation}
where $\delta _{2}\left( r\right) $\ is the generalized two--dimensional $
\delta $--function in flat space. From Eq. (\ref{tensorR}), it follows that
\begin{equation}
R_{12}^{12}:\left\{
\begin{array}{ll}
>0, & \text{if }0<\alpha <1, \\
<0, & \text{if }\alpha >1.
\end{array}
\right.
\end{equation}
In other words, when the defect carries a negative curvature, we have an
excess of the planar angle, which corresponds to an anti--cone. However, if
the defect presents a positive curvature, we have a planar deficit angle,
and the result leads to a cone. For the motion of the particle on a cone,
the geometric potential $V_{s}(r)$, which is a consequence of a
two--dimensional confinement on the surface, is found to be \cite
{PRA.1981.23.1982}
\begin{equation}
V_{s}(\mathbf{r})=-\frac{1}{2M}\left( \mathcal{H}^{2}-\mathcal{K}\right) ,
\label{geometric}
\end{equation}%
where $\mathcal{H}$ is the mean curvature and $\mathcal{K}$ is the Gaussian
curvature of the surface. For the cone ($\alpha <1$), these quantities are
given by \cite{EPL.2007.80.46002}:
\begin{equation}
\mathcal{K}_{cone}=\left( \frac{1-\alpha }{\alpha }\right) \frac{\delta (r)}{
r},  \label{gauss1}
\end{equation}
and
\begin{equation}
\mathcal{H}_{cone}=\frac{\sqrt{1-\alpha ^{2}}}{2\alpha r}.
\end{equation}
In this case, the potential $V_{s}(\mathbf{r})$ reads
\begin{equation}
\left[ V_{s}(\mathbf{r})\right] _{cone}=\frac{1}{2M}\left[ -\frac{(1-\alpha
^{2})}{4\alpha ^{2}r^{2}}+\left( \frac{1-\alpha }{\alpha }\right) \frac{
\delta (r)}{r}\right] .
\end{equation}
However, for the anti--cone ($\alpha >1$), in order to be consistent with
the fact that it has a negative curvature, i.e., the surface takes a saddle
like surface form and the mean curvature is now given by \cite
{JMP.2012.53.122106}
\begin{equation}
\mathcal{H}_{anti-cone}=\frac{\sqrt{\alpha ^{2}-1}}{2\alpha \rho },
\end{equation}
and the geometric potential $V_{s}(\mathbf{r})$ becomes
\begin{equation}
\left[ V_{s}(\mathbf{r})\right] _{anti-cone}=\frac{1}{2M}\left[ +\frac{
(1-\alpha ^{2})}{4\alpha ^{2}r^{2}}+\left( \frac{1-\alpha }{\alpha }\right)
\frac{\delta (r)}{r}\right] .  \label{nu3}
\end{equation}
The magnetic flux tube in the background space described by the metric (\ref
{metric}), which will be our choice for $g_{ij}$, is related to the vector
potential as ($\mathbf{\nabla }\cdot \mathbf{A}=0$, $A_{3}=0$)
\begin{equation}
V\left( \mathbf{r}\right) =0,\qquad QA_{i}=\phi \epsilon _{ij}\frac{r_{j}}{
\alpha r^{2}},  \label{vectorb}
\end{equation}
where $\epsilon _{ij}=-\epsilon _{ji}$ with $\epsilon _{12}=+1$; $\phi =\Phi
/\Phi _{0}$ is the flux parameter with $\Phi _{0}=2\pi /Q$ and the magnetic
field is
\begin{equation}
QB=-\frac{\phi }{\alpha }\frac{\delta (r)}{r}.  \label{vectora}
\end{equation}
In this manner, assuming a solution of the form $\psi _{S}=e^{-iEt}\chi _{S}$
, the Schr\"{o}dinger equation (\ref{hspinless}) is now written as
\begin{equation}
-\frac{1}{2M}\left[ \frac{\partial ^{2}}{\partial r^{2}}+\frac{1}{r}\frac{
\partial }{\partial r}+\frac{1}{\alpha ^{2}r^{2}}\left( \frac{\partial ^{2}}{
\partial \varphi ^{2}}-\frac{2\phi }{i}\frac{\partial }{\partial \varphi }
-\phi ^{2}\right) + \frac{1-\alpha ^{2}}{4\alpha ^{2}r^{2}}-\left( \frac{
1-\alpha }{\alpha }\right) \frac{\delta (r)}{r}\right] \chi _{S}=E\chi _{S}.
\label{schrof}
\end{equation}
We seek eigenfunctions of the form
\begin{equation}
\chi _{S}(r,\varphi )=e^{im\varphi }f_{m}(r),  \label{decompo}
\end{equation}
with $m\in \mathbb{Z}$. Substituting this solution into Eq. (\ref{schrof}),
we obtain for $f(r)$:
\begin{equation}
hf_{m}(r)=k^{2}f_{m}(r),  \label{eigen}
\end{equation}
where $k^{2}=2ME$,
\begin{equation}
h=h_{0}+\left( \frac{1-\alpha }{\alpha }\right) \frac{\delta (r)}{r},
\label{hfull}
\end{equation}
with
\begin{equation}
h_{0}=-\frac{d^{2}}{dr^{2}}-\frac{1}{r}\frac{d}{dr}+\frac{\lambda^{2}
}{r^{2}}  \label{hzero}
\end{equation}
being the Hamiltonian without the $\delta $--function and
\begin{equation}
\lambda^{2}=\frac{4(m+\phi )^{2}- (1-\alpha ^{2})}{4\alpha ^{2}}
\label{angular}
\end{equation}
is the effective angular momentum. The particle, with its motion
confined to the conical surface, is therefore subjected to a generalized
potential $V_{g}(r)$ given by
\begin{equation}
V_{g}\left( r\right) =\frac{\lambda ^{2}}{r^{2}}+\left( \frac{1-\alpha }{
\alpha }\right) \frac{\delta (r)}{r}.  \label{vgn}
\end{equation}
Let us analyze this potential.
For $0< \alpha <1$, the quantity $(1-\alpha^{2})>0$, and in this case
$\lambda^{2}<0$ for $4(m+\phi)^{2}<(1-\alpha ^{2})$,
or $m+\phi =0$.
On the other hand, $(1-\alpha)/\alpha>0$, in such way that we have a
repulsive $\delta$--function.
However, we will see below that, even though the $\delta$--function being
repulsive, the condition $\lambda^{2}<0$ guarantees the existence of
bound states.
For $\alpha >1$, the quantity $(1-\alpha^{2})<0$, consequently we have
$\lambda^{2}>0$ for all values of $m+\phi$, $(1-\alpha)/\alpha>0$,
and the $\delta $--function is now attractive.
In this manner, the attractive $\delta$--function potential guarantees
at least one bound state.
In the Table \ref{tab:tab1} bellow, we summarize the possible physical
scenarios of obtaining $\lambda^{2}>0$ and $\lambda ^{2}<0$ for
$\alpha>1$ and $\alpha <1$.
The case $\alpha =1$ is not of interest here because it implies in a
flat space.
We also summarize the possible physical scenarios of obtaining
scattering and bound states in Table \ref{tab:tab2}, based on the signal
of $(1-\alpha)/\alpha $ in Eq. (\ref{hfull}), for $\alpha \lessgtr 1$.
\begin{table}[h]
\centering
\begin{tabular}{cccc}
\hline
$\alpha$ & $geometry$ & $\lambda^{2}$ & requirement \\ \hline
$>1$ & anti--cone & $>0$ & $\forall$ $(m+\phi)$ \\
$<1$ & cone & $<0$ & $(m+\phi )^{2}<(1-\alpha^{2})/4$ \\ \hline
\end{tabular}
\caption{Summary for the physical scenarios based in $\protect\alpha
\gtrless 1$ for the sign of $\protect\lambda ^{2}$.}
\label{tab:tab1}
\end{table}
\begin{table}[h]
\centering
\begin{tabular}{cccc}
\hline
$\alpha$ & $geometry$ & $(1-\alpha)/\alpha$ & State \\ \hline
$<1$ & cone & $>0$ & Scattering \\
$>1$ & anti--cone & $<0$ & Bound and Scattering \\ \hline
\end{tabular}
\caption{Summary for the physical scenarios based on the signal of $(1-
\protect\alpha )/\protect\alpha $ for $\protect\alpha \lessgtr 1$.}
\label{tab:tab2}
\end{table}
We also summarize the possible physical scenarios of obtaining scattering
and bound states in Table \ref{tab:tab2}, based on the signal of $(1-\alpha
)/\alpha $ in Eq. (\ref{hfull}), for $\alpha \lessgtr 1$.

\section{Particle describing a circular path}

\label{sec:CP}

In this section, we analyze a particularity of the present system, which is
the simple case when a particle is constrained to move in a circle of radius
$r=R$ (for example, bead on a wire ring). In this case, the wave functions
in Eq. (\ref{schrof}) depend only on the azimuthal angle $\varphi $, so that
$\mathbf{\nabla}_{\alpha}\rightarrow (\hat{\mathbf{\varphi}}/\alpha
R)\partial_{\varphi}$. This way, the Schr\"{o}dinger equation yields a
linear differential equation with constant coefficients:
\begin{equation}
\left( \frac{d^{2}}{d\varphi^{2}}+2i\phi \frac{d}{d\varphi}+\mathcal{E}
\right) \chi_{S}=0,  \label{spconst}
\end{equation}
where $\mathcal{E}=2M\alpha^{2}R^{2}E-\phi^{2}+ (1-\alpha^{2})/4$. By
assuming eigenfunctions of the form
\begin{equation}
\chi_{S}(\varphi)=Ae^{im\varphi},  \label{sol1}
\end{equation}
where $A$ is a constant, and substituting it into the Eq. (\ref{spconst}),
one achieves the following solution for the characteristic equation:
\begin{equation}
m=-\phi \pm \sqrt{\phi^{2}+\mathcal{E}}.
\end{equation}
For the wave function $\psi (\varphi )$ to be single--valued, in $
\varphi=2\pi $, the parameter $m$ must be an integer. With this condition,
we obtain discrete values for the energy, namely,
\begin{equation}
E_{m}=\frac{4(m+\phi)^{2} -(1-\alpha^{2})}{8M\alpha^{2}R^{2}},\qquad
m=0,\pm 1,\pm 2,\ldots,  \label{energy1}
\end{equation}
which depends on the mean curvature $\mathcal{H}$. If $\alpha=1$, we fall
into the problem of a charged particle on a circular ring which a long
solenoid passing through it leads to the energy levels for the usual AB
problem \cite{griffiths.QM},
\begin{equation}
E_{m}=\frac{(m+\phi )^{2}}{2MR^{2}}, \qquad m=0,\pm 1,\pm 2,\ldots,
\end{equation}
recovering the lifting of twofold degeneracy of the system due to the
presence of the magnetic flux tube.

\section{Bound state energy and wave function}

\label{sec:SA}

In this section, we obtain the bound state energies and wave functions of
the system. We know, from Ref. \cite{PRD.2012.85.041701}, that the form of
the Hamiltonian (\ref{hfull}) requires a procedure of physical
regularization because of the presence of the $\delta $--function. Before we
proceed further with this approach, it is important to check what are the
criteria revealed by the Hamiltonian (\ref{hfull}) to produce physically
acceptable results.

We commence by observing that when we deal with singular potentials we need
to guarantee that the operator is essentially self-adjoint in order to make
sure that the time evolution is unitary. An operator $O$\ is said to be
essentially self-adjoint if and only if $D(O^{\dagger})=D(O)$\ and $
O^{\dagger}=O$. One observes that even if such operator is Hermitian, i.e., $
O^{\dagger}=O$, its domain could be different from its adjoint. Roughly
speaking, the self-adjoint extension approach consists, essentially, in
extending the domain $D(O)$\ in order to match $D(O^{\dagger})$. In the
present case, if $g\in C_{0}^{\infty}( \mathbb{R}^{2})$, with $C_{0}^{\infty }( \mathbb{R}^{2})$
denoting the set of functions that is differentiable for all degrees of
differentiation, and $g(0)=0$, $h$ should coincide with $h_{0}$, in such way
that $hg=h_{0}g$ \cite{Book.2004.Albeverio,JMP.1980.21.840}. Thus it is
reasonable to interpret $h$\ as an extension of $h_{0}$, or more precisely,
as a self--adjoint extension of $h_{0}|_{C_{0}^{\infty}(\mathbb{R }
^{2}\setminus \{0\})}$ \cite{crll.1987.380.87,JMP.1998.39.47,LMP.1998.43.43}. Using the unitary operator $U:L^{2}(\mathbb{R}^{+},rdr)\rightarrow L^{2}(
\mathbb{R}^{+},dr)$, given by $(Ug)(r)=r^{1/2}g(r)$, the operator $h_{0}$
can be written as
\begin{equation}
\bar{h}_{0}=Uh_{0}U^{-1}= -\frac{d^{2}}{dr^{2}}+\frac{1}{r^{2}}\left(
\lambda^{2}-\frac{1}{4}\right) .
\end{equation}
As a result, it is well--known \cite{Reed.2.1975} that the symmetric radial
operator $\bar{h}_{0}$ is essentially self--adjoint for $\lambda^{2}\geq 1$
. On the other hand, if $\lambda^{2}<1$, it is not essentially
self--adjoint, admitting an one--parameter family of self--adjoint
extensions. To characterize this one-parameter family of $h_{0}$, we will
use the approach of Ref. \cite{CMP.1991.139.103}, which is based on boundary
conditions. Basically, the boundary condition is a match of the logarithmic
derivatives of the zero--energy solutions for Eq. (\ref{eigen}) and the
solutions for the problem $h_{0}$ plus self--adjoint extension.

In this manner, we solve the problem without the $\delta $--function
potential and then we find the boundary condition by invoking the
self--adjointness of $h_{0}$. For this, we must solve the eigenvalue
equation
\begin{equation}
h_{0}f_{\varrho}=k^{2}f_{\varrho},  \label{ideal}
\end{equation}
plus self--adjoint extensions. Here, the label $\varrho $ is the
self--adjoint extension parameter, which is related to the behavior of the
wave function at the origin. In order for $h_{0}$ to be a self--adjoint
operator, its domain has to be extended by the deficiency subspace, which is
given by the solutions of the eigenvalue equation
\begin{equation}
h_{0}^{\dagger}f_{\pm}=\pm if_{\pm}.  \label{eigendefs}
\end{equation}
In the next sections, we will use the present approach to determine the
energy spectrum for a particle lying on an anti--cone and on a cone.

\subsection{Quantum dynamics on an anti--cone}

\label{sec:AC}

According to the Table \ref{tab:tab1}, $\alpha >1$ implies $\lambda^{2}>0$
and the particle lies on an anti--cone. In this case, by solving Eq. (\ref
{eigendefs}), the only square integrable functions which are solutions are
the modified Bessel functions of second kind
\begin{equation}
f_{\pm}=K_{|\lambda |}(\sqrt{\mp i}r),
\end{equation}
with $\Im \sqrt{\pm i}>0$. These functions are square integrable only in the
range $|\lambda |<1$ and, as stated above, in this range $h_{0}$ is not
self--adjoint. The dimension of such deficiency subspace is found to be $
(n_{+},n_{-})=(1,1)$. Thus, the domain $\mathcal{D}(h_{0,\varrho})$ of the
extended operator $h_{0,\varrho}$ in $L^{2}(\mathbb{R}^{+},rdr)$ is given by
the set of functions \cite{Reed.2.1975}
\begin{equation}
f_{\varrho}(r)=f_{m}(r)+C\left[ K_{|\lambda |}(\sqrt{-i}r)+e^{i\varrho
}K_{|\lambda |}(\sqrt{i}r)\right] ,  \label{domain}
\end{equation}
where $f_{m}(r)$ is the regular wave function with $f_{m}(0)=df_{m}(0)/dr=0$
and the parameter $\varrho \in \lbrack 0,2\pi )$ represents a choice for the
boundary condition. For each different $\varrho $, we have a possible domain
for $h_{0}$ and the physical situation will determine the value of $\varrho $
\cite{AoP.2010.325.2529,AoP.2008.323.3150,PRD.2012.85.041701,
AoP.2013.339.510,PLB.2013.719.467,EPJC.2014.74.3187, EPJC.2014.74.3112} .
Thus, in order to find a fitting for $\varrho $ compatible with the physical
situation, we require a physical regularization for the $\delta $
--function. This is accomplished by replacing \cite{PRL.1990.64.503}
\begin{equation}
\frac{\delta (r)}{r}\rightarrow \frac{\delta (r-a)}{a},  \label{deltareg}
\end{equation}
with $a$ representing the nucleus of a real physical system.

In order to find the energy levels, we need at first to determine a value
for $\varrho $ compatible with the physics imposed by the regularized $
\delta $--function in Eq. (\ref{deltareg}). Following \cite{CMP.1991.139.103}, we consider the zero--energy solutions $f_{0}$ and $f_{\varrho ,0}$ for $h$
with the regularization in Eq. (\ref{deltareg}) and $h_{0}$, respectively,
i.e.,
\begin{equation}
\left[ -\frac{d^{2}}{dr^{2}}-\frac{1}{r}\frac{d}{dr}+\frac{\lambda^{2}}{
r^{2}}+\left( \frac{1-\alpha}{\alpha}\right) \frac{\delta (r-a)}{a}\right]
f_{0}=0,  \label{statictrue}
\end{equation}
\begin{equation}
\left[ -\frac{d^{2}}{dr^{2}}-\frac{1}{r}\frac{d}{dr}+\frac{\lambda^{2}}{
r^{2}}\right] f_{\varrho ,0}=0,  \label{rhostatic}
\end{equation}
and the value of $\varrho $ is determined by the boundary condition
\begin{equation}
\lim_{a\rightarrow 0^{+}}\frac{r}{f_{0}(r)}\frac{df_{0}(r)}{dr}\Big|
_{r=a}=\lim_{a\rightarrow 0^{+}}\frac{r}{f_{\varrho ,0}(r)}\frac{df_{\varrho
,0}(r)}{dr}\Big|_{r=a}.  \label{logder}
\end{equation}
The left--hand side of Eq. (\ref{logder}) is determined by integrating Eq. (
\ref{statictrue}) from $0$ to $a$ using the property that $f_{0}$ must be
finite at the origin, yielding
\begin{equation}
\lim_{a\rightarrow 0^{+}}\frac{r}{f_{0}(r)}\frac{df_{0}(r)}{dr}\Big|_{r=a}=
\frac{1-\alpha}{\alpha}.  \label{nrs}
\end{equation}
In order to find the right--hand side of Eq. (\ref{logder}), we use the
relation (\ref{domain}) and write $K_{\lambda}(z)$ in terms of $I_{\lambda}(z)$ as
\begin{equation}
K_{\lambda}(z)=\frac{\pi}{2\sin (\pi \lambda )}\left[ I_{-\lambda}(z)-I_{\lambda}(z) \right]
.  \label{kni}
\end{equation}
Using the series expansion for $I_{\lambda}$,
\begin{equation}
I_{\lambda}=\left( \frac{z}{2}\right)^{\lambda}\sum_{k=0}^{\infty}\frac{
(z^{2}/4)^{k}}{k!\Gamma (\lambda +k+1)}  \label{series}
\end{equation}
into the Eq. (\ref{kni}), we can get the following asymptotic form ($
z\rightarrow 0$):
\begin{equation}
K_{\lambda}(z)\sim \frac{\pi}{2\sin (\pi \lambda )}\left[ \frac{z^{-\lambda}}{2^{-\lambda
}\Gamma (1-\lambda )}-\frac{z^{\lambda}}{2^{\lambda}\Gamma (1+\lambda )}\right] .
\label{besselasympt}
\end{equation}
Using this result, one finds
\begin{equation}
\lim_{a\rightarrow 0^{+}}\frac{r}{f_{\varrho ,0}(r)}\frac{df_{\varrho ,0}(r)
}{dr}\Big|_{r=a}=\lim_{a\rightarrow 0^{+}}\frac{1}{\Theta_{\varrho}(r)}
\frac{d\Theta_{\varrho}(r)}{dr}\Big|_{r=a},  \label{dright}
\end{equation}
where
\begin{equation}
\Theta_{\varrho}(r)=\left[ \frac{\left( \sqrt{-i}r\right)^{-|\lambda |}}{
2^{-|\lambda |}\Gamma \left( 1-|\lambda |\right)}-\frac{\left( \sqrt{-i}
r\right)^{|\lambda |}}{2^{|\lambda |}\Gamma \left( 1+|\lambda |\right)}
\right] +e^{i\varrho}\left[ \frac{\left( \sqrt{i}r\right)^{-|\lambda |}}{
2^{-|\lambda |}\Gamma \left( 1-|\lambda |\right)}-\frac{\left( \sqrt{i}
r\right)^{|\lambda |}}{2^{|\lambda |}\Gamma \left( 1+|\lambda |\right)}
\right] .
\end{equation}
By inserting Eqs. (\ref{nrs}) and (\ref{dright}) into Eq. (\ref{logder}), we
obtain
\begin{equation}
\lim_{a\rightarrow 0^{+}}\frac{1}{\Theta_{\varrho}(r)}\frac{d\Theta
_{\varrho}(r)}{dr}\Big|_{r=a}=\frac{1-\alpha}{\alpha}.  \label{saepapprox}
\end{equation}
This result gives us the parameter $\varrho $ compatible with the physics
imposed by the problem. In other words, it gives the correct behavior for
the wave function when $r\rightarrow 0^{+}$.

We are now in position to determine the bound states for $h$. So, we write
Eq.(\ref{ideal}) for the bound state. In the present system, the energy of a
bound state has to be negative so that $k$ is a pure imaginary number, $
k=i\kappa $ with $\kappa=\sqrt{-2ME}$ and $E<0$. Then, by exchanging $
k\rightarrow i\kappa$, we have
\begin{equation}
\left[ \frac{d^{2}}{dr^{2}}+\frac{1}{r}\frac{d}{dr} -\left( \frac{\lambda^{2}
}{r^{2}}+\kappa^{2}\right) \right] f_{\varrho}(r)=0.  \label{eigenvalue}
\end{equation}
The solution of Eq. (\ref{eigenvalue}) is given by the modified Bessel
function
\begin{equation}
f_{\varrho}(r)=K_{|\lambda|}\left( r\sqrt{-2ME}\right) .  \label{sver}
\end{equation}
Notice that the solution (\ref{sver}) is of the form (\ref{domain}) for some
$\varrho$ selected from the physics of the problem. Then, by substituting
Eq. (\ref{sver}) into Eq. (\ref{domain}) and using Eq. (\ref{besselasympt}),
we compute the quantity
\begin{equation}
\lim_{a\rightarrow 0^{+}}a\frac{{\dot{f}_{\varrho}}\left( r\right)}{{\
f_{\varrho}}\left( r\right)}\Big|_{r=a}.
\end{equation}
A straightforward calculation yields
\begin{equation}
\frac{|\lambda |\left[ a^{2|\lambda |}\Gamma \left( 1-|\lambda |\right)
(-ME)^{|\lambda |}+2^{\left\vert j\right\vert}\Gamma \left( 1+|\lambda
|\right) \right]}{\alpha a^{2\left\vert j\right\vert}\Gamma \left(
1-|\lambda |\right) (-ME)^{|\lambda |}-2^{|\lambda |}\Gamma \left(
1+|\lambda |\right)}=\frac{1-\alpha}{\alpha}.  \label{derfe}
\end{equation}
Solving the above equation for $E$, we find the sought energy spectrum
\begin{equation}
E=-\frac{2}{Ma^{2}}\left[ \left( \frac{1-\alpha +\alpha |\lambda |}{1-\alpha
-\alpha |\lambda |}\right) \frac{\Gamma (1+|\lambda |)}{\Gamma (1-|\lambda
|) }\right]^{\frac{1}{|\lambda |}}.  \label{energy_KS}
\end{equation}
Notice that there is no arbitrary parameter in the above equation. Moreover,
in order to ensure that the energy is a real number, we must have
\begin{equation}
\left( \frac{1-\alpha +\alpha |\lambda |}{1-\alpha -\alpha |\lambda |}
\right) \frac{\Gamma (1+|\lambda |)}{\Gamma (1-|\lambda |)}>0.
\end{equation}
This inequality is satisfied if $\left\vert 1-\alpha \right\vert \geq 1\geq
|\lambda |$ and due to the fact taht $|\lambda |<1$, it is sufficient to
consider $\left\vert 1-\alpha \right\vert \geq 1$. As shown in Table \ref
{tab:tab1}, a necessary condition for a $\delta $--function to generate an
attractive potential, which is able to support bound states, is that the
coupling constant $(1-\alpha )/\alpha $ must be negative.

\subsection{Quantum dynamics on a cone}

As mentioned above, the only possibility to generate a cone is $\alpha<1$,
implying $\lambda^{2}<0$ only if $(m+\phi)^{2}<(1-\alpha^{2})$ (see Table
\ref{tab:tab1}). In this case, the Schr\"{o}dinger equation reads (with $
\lambda^{2}<0$) as
\begin{equation}
\tilde{h}_{0}f_{\tilde{\varrho}}=k^{2}f_{\tilde{\varrho}},  \label{nws}
\end{equation}
plus self--adjoint extensions, with
\begin{equation}
\tilde{h}_{0}=-\frac{d^{2}}{dr^{2}}-\frac{1}{r}\frac{d}{dr} -\frac{
\lambda^{2}}{r^{2}},  \label{hnws}
\end{equation}
whose solution outside the origin is now given by
\begin{equation}
f_{\tilde{\varrho}}(r)=K_{i|\lambda|}\left( r\sqrt{-2mE}\right) ,  \label{sol2}
\end{equation}
which is the modified Bessel function of purely imaginary order \cite
{Abramowitz1964}.

Next, following the same recipe of the previous section, we must solve the
following equations:
\begin{equation}
\left[ -\frac{d^{2}}{dr^{2}}-\frac{1}{r}\frac{d}{dr}-\frac{\lambda^{2}}{
r^{2}}+\left( \frac{1-\alpha}{\alpha}\right) \frac{\delta (r-a)}{a}\right]
f_{0}=0,  \label{prba}
\end{equation}
\begin{equation}
\left[ -\frac{d^{2}}{dr^{2}}-\frac{1}{r}\frac{d}{dr}-\frac{\lambda^{2}}{
r^{2}}\right] f_{\tilde{\varrho},0}=0.  \label{prbb}
\end{equation}

An asymptotic expansion for the modified Bessel function of the pure
imaginary order is obtained by replacing $\lambda$ by $i\lambda $ in Eq. (\ref
{besselasympt}) and by writing
\begin{equation}
\Gamma (1\pm i\lambda )=\left( \frac{\pi \lambda }{\sinh \pi \lambda }\right) ^{\frac{1}{
2}}e^{\pm i\gamma _{\lambda }},
\end{equation}
where $\gamma _{\lambda }$ is the Coulomb phase shift \cite{Taylor2006}. In this
manner, we get at
\begin{equation}
K_{i|\lambda |}(x)\sim -\left( \frac{\pi }{|\lambda |\sinh \pi |\lambda |}
\right) ^{\frac{1}{2}}\sin \left[ |\lambda |\ln \left( \frac{x}{2}\right)
-\gamma _{|\lambda |}\right] .  \label{nu}
\end{equation}
By using the boundary condition (\ref{logder}), we obtain
\begin{equation}
\lim_{a\rightarrow 0^{+}}a\frac{\dot{f}_{\tilde{\varrho},0}}{f_{\tilde{
\varrho},0}}\Big|_{r=a}=\lim_{a\rightarrow 0^{+}}\frac{\dot{\xi}_{\tilde{
\varrho}}(r)}{\xi _{\tilde{\varrho}}(r)}\Big|_{r=a},  \label{bound1}
\end{equation}
where
\begin{equation}
\xi _{\tilde{\varrho}}(r)=\sin \left[ |\lambda |\ln \left( \frac{1}{2}\sqrt{
-2Mi}r\right) +\gamma _{|\lambda |}\right] +e^{i\tilde{\varrho}}\sin \left[
|\lambda |\ln \left( \frac{1}{2}\sqrt{+2Mi}r\right) +\gamma _{|\lambda |}
\right] .  \label{F}
\end{equation}
Integration of Eq. (\ref{prba}) from $0$ to $a$ provides the left--hand side
of Eq. (\ref{logder}). The result of this operation is given in Eq. (\ref
{nrs}). So, from Eqs. (\ref{logder}), (\ref{bound1}) and (\ref{nrs}), we
arrive at
\begin{equation}
\lim_{a\rightarrow 0^{+}}\frac{\dot{\xi}_{\tilde{\varrho}}(r)}{\xi _{\tilde{
\varrho}}(r)}\Big|_{r=a}=\frac{1-\alpha }{\alpha }.  \label{firstfit}
\end{equation}
In order to find the bound states of $\tilde{h}_{0}$, we use Eq. (\ref{sol2}
) together with Eq. (\ref{nu}), which provides
\begin{equation}
\xi _{\tilde{\varrho}}(r=a)=\sin \left[ |\lambda |\ln \left( \frac{a}{2}
\sqrt{-2ME}\right) +\gamma _{|\lambda |}\right] ,
\end{equation}
and
\begin{equation}
\dot{\xi}_{\tilde{\varrho}}(r=a)=\frac{|\lambda |}{a}\cos \left[ |\lambda
|\ln \left( \frac{a}{2}\sqrt{-2ME}\right) +\gamma _{|\lambda |}\right] .
\end{equation}
By replacing the above expressions in Eq. (\ref{firstfit}), we get
\begin{equation}
|\lambda |\cot \left[ |\lambda |\ln \left( \frac{a}{2}\sqrt{-2ME}\right)
+\gamma _{|\lambda |}\right] =\frac{1-\alpha }{\alpha }.
\end{equation}
Solving this equation for $E$, we find
\begin{equation}
E=-\frac{2}{Ma^{2}}\exp \left[ \frac{2}{\alpha \sqrt{(m+\phi )^{2}-(1-\alpha
^{2})/4}}\cot ^{-1}\left( \frac{(1-\alpha )}{\alpha \sqrt{(m+\phi
)^{2}-(1-\alpha ^{2})/4}}-\gamma _{|\lambda |}\right) \right] .
\label{novel}
\end{equation}
Equation (\ref{novel}) reveals that, while the mean curvature $\mathcal{H}$
contributes attractively, the Gaussian curvature $\mathcal{K}$ contributes
with a repulsive $\delta $--function potential. Thus, Eq. (\ref{angular})
implies that the only allowable value for the angular momentum is $m+\phi $,
meaning that we have a single bound state. In other words, when the $\delta $
--function potential is repulsive ($\alpha <1$) an attractive effective
potential potential assures one bound state ($m+\phi =0$).

\section{Conclusions}

\label{sec:conclusions}

In this paper, we have studied the dynamics of a spinless charged particle
which moves bounded to a 2D surface immersed in an 3D Euclidean space and in
the presence of the Aharonov--Bohm potential. In other words, we have solved
a spinless Aharonov--Bohm--like problem in curved space. The motion of the
particle is decomposed into two, being one on the surface and the other in a
normal direction in relation to the surface. The dynamics on the surface is
governed by the Schr\"{o}dinger equation (Eq. (\ref{hspinless})) coupled to
the potential vector, while on the normal direction the dynamics is given by
Eq. (\ref{q3}), which is the equation for an infinity curved quantum well.
The surface mean and Gaussian curvatures enter in the Schr\"{o}dinger
equation as a scalar potential, namely, as a geometric potential. We chose a
conical surface which is defined by the metric endowed in the line element in Eq. (
\ref{metric}).

A particularity of this system is that the isotropy of space is broken which
means that the velocity operators do not commute with each other. Such a
feature is usually due to the presence of a magnetic field. However, in the
context, we also have effects of the geometry of the surface since the
connection of the space plays an important role. To describe the full
dynamics of the system, we have analyzed three situations. First, we have
found the energy eigenvalues and wave functions for the simple case of the
particle describing a circular path around the solenoid. In the other two
cases, we have considered the dynamics in the full space, including the $r=0$
region. For those cases, the geometry of the system dictated by the line
element in Eq. (\ref{metric}) establishes that the motion of a particle can occurs
on the surface of a cone (for $\alpha <1$) or on the surface of an
anti--cone (for $\alpha >1$). Expressions for the bound state energies and
wave functions were obtained for both cases.

\section*{Acknowledgments}

We acknowledge some suggestions made by the anonymous referees in order to improve
the present work. This work was supported by the CNPq, Brazil, Grants Nos. 482015/2013--6
(Universal), 476267/2013--7 (Universal), 460404/2014-8 (Universal), 306068/2013--3 (PQ), 206224/2014-1 (PDE) and FAPEMA, Brazil, Grants Nos. 00845/13 (Universal), 01852/14 (PRONEM).

\bibliographystyle{model1a-num-names}

\end{document}